# Terahertz radiation generated by shell electrons in the bubble regime via the interaction between an intense laser and underdense plasma

J. F. Qu [1], X. F. Li [2,3], X. Y. Liu [1], P. Liu [1], Y, J. Song [1], Z. Fu [1], Q. Yu [4] and Q. Kong [1*]

[1] Key Laboratory of Nuclear Physics and Ion-beam Application (MOE), Institute of Modern Physics, Department of Nuclear Science and Technology, Fudan University, Shanghai, China 200433

[2] Key Laboratory for Laser Plasmas (MoE), School of Physics and Astronomy, Shanghai Jiao Tong University, Shanghai 200240, China

[3] Collaborative Innovation Center of IFSA (CICIFSA), Shanghai Jiao Tong University, Shanghai 200240, China

[4] Shanghai Institute of Optics and Fine Mechanics, Chinese Academy of Sciences, Shanghai 201800, China

## Abstract

Backward terahertz radiation can be produced by a high-intensity laser normally incident upon an underdense plasma. It is found that terahertz radiation is generated by electrons refluxing along the bubble shell. These shell electrons have similar dynamic trajectories and emit backward radiations to vacuum. This scheme has been proved through electron dynamic calculations as well as by using an ionic sphere model. In addition, the bubble shape is found to influence the radiation frequency, and this scheme can be implemented in both uniform and up-ramp density gradient plasma targets. The terahertz radiation may be used for diagnosing the electron bubble shape in the interaction between an intense laser and plasma. All results are presented via 2.5 dimensional particle-in-cell simulations.

## I. INTRODUCTION

The terahertz (THz) range intersects the infrared and microwave regions in the frequency spectrum and has a wide application prospect ranging from macroscopic electronics to microcosmic photonics [1], which means that it is essential to identify high-quality sources for THz frequencies. With the emergence of the femtosecond laser [2–6], and the intensity of laser pulses being high, plasma has become an optimal choice, mainly relying on the advantage of unlimited threshold. So far, many schemes [7-10] and THz generation mechanisms [11-20] based on laser–plasma interactions have been proposed. Sheng et al. [19] found that a nonrelativistic laser interacting with the density gradient of an inhomogeneous plasma can produce THz pulses through linear mode conversion [7], which has been proven experimentally [21]. When the laser intensity becomes high, owing to the relativistic effect and the tendency of the wakefield to break transversely, the linear mode conversion model cannot describe the wake emission well, and the induced THz frequency spectrum becomes broad and irregular [20]. This phenomenon was also found by Hu et al. [22]; the THz radiation was found to detach in a highly nonlinear bubble wakefield via the underdense plasma, adding to the external DC magnetic field. All these studies mentioned that the emitted pulses are regular only for a few cycles at the beginning; however, this phenomenon was not analyzed, and the mechanism involved remains unsolved.

In this study, we detect the backward THz radiation when a relativistic laser normally incident upon an underdense plasma. As is well known, when an intense focused laser pulse enters an underdense plasma, the plasma electrons are expelled radially outward leaving behind a channel of ions, because of the radiation pressure of the intense laser. These electrons, which are expelled and blown out, form a narrow sheath just outside the ion channel, and they will eventually be pulled back by the space-charge field, called bubble structure [23-24]. In bubble structure, some of electrons can be trapped inside the bubble, most others are not injected into the bubble. Present research also show that the electron not injecting into the bubble eject from plasma at an angle of 30º-60º for the laser propagation axis and emit forward the THz radiation [25-26]. Here, the bubble shell electrons movement backward are considered. That is, in the bubble structure, the bubble shell is composed of a large number of electrons moving backward at any time. These shell electrons reflux with those having a similar dynamic trajectory. We believe that the backward THz radiation is caused by these backflow electrons. The entire THz generation process is verified using 2.5-dimensional particle-in-cell (PIC) simulations based on the electromagnetic relativistic code EPOCH [27] and a spherical bubble radiation model is proposed. THz radiation can occur not only in uniform plasma but also in plasma with an up-ramp density gradient. The THz radiation frequency is related to the bubble radius for different parameters of laser intensity and plasma density.

## II. THEORY AND SIMULATION RESULTS

In the present work, a Gaussian focused laser pulse is used with a linear P-polarized wave propagating along the x-direction:

$$E = E_0 \frac{w(x)}{w_0} \exp(-\frac{y^2+z^2}{w^2(x)}) \exp(-\frac{(t-\tau)^2}{(0.5\tau)^2}) \cos(\phi), \tag{1}$$

where $w(x) = w_0[1+(2x^2/kw_0^2)]^{0.5}$, $\tau$ is the pulse duration, $w_0$ is the laser beam waist, $\lambda_0 = 1.06\mu m$ is the wavelength, and all the lengths are normalized by $\lambda_0$. For simplicity, a dimensionless variable $a_0 = eE_0/m_e\omega c$ is used to represent the laser intensity, and the relationship between $a_0$ and I is given by $I = 1.37\times 10^{18} a_0^2/\lambda_0^2 \ (W/cm^2 \bullet \mu m)$, where $e$ and $m_e$ are the electron charge and electron mass, respectively. A laser beam with $a_0 = 10, w_0 = 10$ and $\tau = 21\text{fs}$ is employed as a typical case for THz radiation generation. A simulation box with a size of $450\times 300$ is set up with a mesh size of dx = 1/32 and dy = 5/32, and each cell is filled with 8 superparticles. The plasma region has an area of $200\times 120$ with an initial density $n_0 = 0.016 n_c$, where $n_c$ and $n_0$ are the critical and initial plasma density, respectively.

Under the above parameters, the laser pulses are intense enough to break the plasma wave and form a cavity behind the laser pulses. The electron bubble exhibits a spherical shape, which can be seen in the electron density snapshot at 810 fs in Fig. 1(a). In the results of the PIC simulation, an electromagnetic (E-M) wave is seen to be emitted backward from the left boundary of the plasma. To describe the wave clearly, the magnetic field snapshot at 900 fs is plotted in Fig. 1(b) in the space scale (in which the entire region is located in vacuum). It is obvious that there is a periodic magnetic field structure in the vacuum behind the plasma, where the blue color indicates a negative value and the red color indicates a positive value. This backward magnetic field oscillates with a much greater wavelength than that of the laser, and it propagates along the center axis of the laser to both sides, as shown in Fig. 1(b).

To confirm the time characteristic of the backward E-M wave in vacuum, a point are selected ($x = 180, y = 170$) that is located in the area of energy concentration in vacuum. The temporal magnetic field waveform from 550 to 900 fs is shown in Fig. 1(c). Initially, from 550 fs to 695 fs, when the laser is incident on the plasma, the bubble is not fully formed. This is due to the computational plasma boundary effect. Then, from 850 to 900 fs, when the laser is injected into the plasma, the THz wave cannot propagate through to vacuum due to the plasma absorption effect, and noise is created. The corresponding Fourier spectrum of the frequency emission is depicted in Fig. 1(d) for the period from 700 to 850 fs. There is one peak value (32.8 THz) in Fig.

1(d).

The E-M wave radiation spectrum can be identified from the aspect of time. Why the THz radiation is emitted backwards into vacuum. We think that this is caused during the reflux of shell electrons. In bubble structure, left ions are surrounded by the shell electrons moving backward. Meanwhile, the bubble shell electrons have three main features [28]: (i) the shell electrons consist of two-beam electrons due to the laser ponderomotive force; (ii) the shell electrons are continuously complementary; (iii) the shell electrons have similar trajectories and velocity around the cavity. We know that relativistic particles in a circular motion emit a electromagnetic radiation. Therefore, when a large number of shell electrons move backward with similar movement trajectories and velocity and emit similar electromagnetic radiation, these electromagnetic radiation are similar in direction and are backwards.

To further verify the above explanation, a simple ionic sphere model is established. Hence, it is assumed that the bubble shell electrons move around the ionic sphere at a uniform circular velocity. According to the synchrotron radiation theory [29], relativistic particles in a circular motion emit a fundamental frequency equal to the angular frequency $\omega_0 = v/R$ and radiate a broadband spectrum up to the critical frequency $\omega_c = (v/R)\gamma^3$, where $v$ is the electron circular motion velocity, $R$ is the radius of curvature, and $\gamma$ is the Lorentz factor of an electron. Lu et al [30] have proposed that the maximum blowout radius should match the laser spot size and plasma density, $k_p R = k_p w_0 \sim 2\sqrt{a_0}$ in the sphere bubble. By using the above parameter and the ionic sphere model, it was calculated that the spherical bubble shell electrons radiate a base frequency of about 30.86 THz and a critical frequency of about 246.88 THz, which is lower than the laser frequency. This result is consistent with the peak value shown in Fig. 1(d). Here, $v$ and $\gamma$ are replaced by the numerical values of the average electron velocity and gamma (at different times) shown in Fig. 2(a).

In order to prove the above model through numerical simulation, some bubble sheath electrons near the middle part of the bubble (in Fig. 1(a)) are selected. Fig. 2(b) shows six typical bubble sheath electron motion trajectories in laboratory coordinates from 877 to 904 fs. The bubble has an almost spherical shape. The sheath electrons along the spherical bubble layer have similar movement trajectories, consistent with the features of bubble shell electrons. To understand the dynamic characteristics of these electrons, a typical electron is chosen (marked by a black arrow in Fig. 2(b)). The evolution of the electron longitudinal velocity is plotted in Fig. 2(c), and we use two blue dashed lines to represent the time interval during which the electron moves along the bubble sheath. From Fig. 2(c), it can be seen that the electron velocity is mostly less than zero, which means it will move backward and emit radiation. During the marked interval, the electron velocity approaches the speed of light. A relativistic charged particle radiation function was derived by Jackson [29]. The radiation power per unit frequency emitted by a single electron is given by

$$\frac{dP}{d\omega} = \frac{\sqrt{3}}{4\pi\varepsilon_0} \frac{e^2\gamma}{c} \frac{\omega}{\omega_c} \int_{\frac{\omega}{\omega_c}}^{\infty} K_{5/3}(\xi) d\xi, \qquad (2)$$

where $\omega_c = 3\gamma^3 c/2\rho$ is the critical frequency, $e$ is the electron charge, $\gamma$ is the Lorentz factor, $\varepsilon_0$ is the vacuum permittivity constant, $c$ is the speed of light in vacuum, $\rho$ is the instantaneous radius of curvature, and $K_{5/3}(\xi)$ is the modified Bessel function of the second kind. The radiation power per unit frequency produced by an electron for a given period of time is $dP/d\omega = \sum_t dP_t/d\omega$. Fig. 2(d) shows the electron radiation power spectrum corresponding to Fig. 2(c) from 877 to 904 fs. The peak value of the power spectrum frequency is located at 33.2 THz, which is consistent with one of the peak values of the Fourier transformation spectrum shown in Fig. 1(d) and approaches the value calculated by the ionic sphere model. The bubble shell consists of a large number of electrons with similar motion trajectories. The detected radiation frequency in the vacuum is sum over all the electrons emitted frequency.

The ionic sphere model and the electron dynamics theory are mainly based on a spherical bubble; however, an ellipsoid bubble is more common. Li et al [31] suggested that the ellipsoid bubble exists generally and proposed describing the bubble shape with the aspect ratio $\eta = R_\parallel/R_\perp$, where $R_\parallel$ is the longitudinal radius and $R_\perp$ is the transverse radius. For the bubble in Fig. 1(a), $\eta = 1$. In an ellipsoid bubble, the electron trajectory can be expressed as $\frac{x^2}{a^2} + \frac{y^2}{\eta^2 a^2} = 1$, where $a$ is the transverse radius. The corresponding radius of curvature is given by

$$\rho = \left|\frac{(a^4 y^2 + a^4 \eta^4 x^2)^{1.5}}{a^8 \eta^4}\right| = \left|\frac{(a^2 - x^2 - y^2 + a^2\eta^2)^{1.5}}{-\eta a^2}\right|. \qquad (3)$$

When the elliptical centrifuge rate is not big, $\eta$ approaches 1, $x^2 + y^2 \approx a^2$, and $\rho \propto a\eta^2$; that is, $\omega \propto c/\eta^2$. In order to verify the model, the numerical simulation results are shown in Fig. 3, which shows how the bubble shape affects the THz radiation peak value in an ellipsoid bubble. The longitudinal radius ($R_\parallel$), the transverse radius ($R_\perp$), and the emitted THz for different plasma densities are shown in Fig. 3(a). According to the work [24], which was based on a spherical bubble, the bubble radius can be estimated from the balance between the Lorentz force of the sheath electrons and the laser ponderomotive force. When the laser beam waist and intensity are kept constant, the transverse radius is almost invariant. However, the Lorentz force is enhanced due to the increase in particle amount and the small longitudinal radius. In addition, with the increase in plasma density, the THz peak value tends to become high. From the ellipsoid model, we know that $\omega \propto 1/\eta^2$. Fig. 3(b) shows the linear relation between the aspect ratio $1/\eta^2$ and the emitted THz peak value obtained through numerical simulation. Here, the laser intensity is $a_0 = 15$, which is identical to that of the ellipsoid model. In addition to the plasma density, the laser intensity can also alter the bubble shape. Fig. 3(c) shows the relationship between the bubble radius, the emitted THz peak value, and the laser intensity. In addition, the longitudinal radius increases with increase in the laser intensity. Unlike that in the case of increase in plasma density, with an increase in laser intensity, the laser ponderomotive force increases (with a plasma density $n_0 = 0.0175n_c$), the particle number remains constant, and the longitudinal radius of the bubble increases. Meanwhile, the radiation frequency of the bubble sheath electron that is generated decreases, as shown in Fig. 3(c). From Fig. 3(d), it is seen that the frequency is consistent with $1/\eta^2$, in accordance with the ellipsoid model. In addition, from the above simulation, it is seen that although the parameters are different (such as different laser intensities and plasma densities), the THz radiation emitted by the bubble sheath electrons is inversely proportional to the aspect ratio $\eta^2$. According to these results, we can diagnose the bubble shape and geometry by measuring the radiation frequency in vacuum.

    In addition to being generated for a uniform density target, THz radiation is also generated in plasma with an up-ramp density gradient. Here, an example is given for a laser pulse with $a_0 = 10, w_0 = 10$, $\tau = 21\text{fs}$, and $\lambda_0 = 1.06\mu m$ propagating in a plasma with an initial density increasing from $n_0 = 0$ to $n_0 = 0.016n_c$ within $200 \le x \le 400$. Figs. 4(a) and 4(b) show the electron density and magnetic field snapshots at 1380 fs, using the same simulation box as that in Fig. 1 with a mesh size $dx = 0.03125, dy = 5dx$, respectively. It can be seen that three electron bubbles are formed in the plasma and that backward electromagnetic waves are generated. Because the physics phenomenon is more distinct in the first bubble, we consider only the first bubble in this paper. We select a point ($x = 180, y = 120$) in vacuum located in the area of energy concentration. The temporal magnetic field waveform at this point is shown in Fig. 4(c). The one-cycle pulse can continue for hundreds of femtoseconds, which is qualitatively

consistent with the period and wavelength shown in Fig. 4(b), in the THz frequency range. The corresponding Fourier spectrum of the low-frequency emission, which is shown in Fig. 4(d), also proves this. The spectral shape shows that the pulse energy is mainly concentrated in the frequency range of 1–10 THz, with a single peak frequency of 7 THz.

### III. CONCLUSIONS

In conclusion, we have confirmed that THz radiation can be generated in the bubble regime when an intense laser is normally incident upon a uniform plasma. The bubble sheath consists of a large number of electrons, and these electrons have similar backward motion trajectories. By using an ionic sphere model and by analyzing the electron dynamics of radiation, we confirmed that backward THz radiation is generated by the sheath electrons. The bubble shape, which depends on the laser intensity and plasma density, affects the THz radiation frequency. Our method might prove useful for studying the laser wakefield and bubble shape. In addition, this scheme is also suitable for studying plasma with an up-ramp density gradient.

### ACKNOWLEDGMENTS

This work was supported by NSFC (No.11175048).

**Figure Caption**

**Figure 1**

Snapshots of (a) the electron density at 810 fs and (b) the magnetic field at 900 fs when a relativistic laser ( $a_0 = 10$, $w_0 = 10$, and $\tau=21\text{fs}$ ) interacts with a uniform plasma with $n_0 = 0.016 n_c$ . (c) Temporal magnetic waveform from 550 to 900 fs (black line) and (d) the corresponding Fourier spectrum of the backward radiation for uniform plasma from 695 to 850 fs (blue line).

**Figure 2**

Bubble shell electrons average velocity and gamma in different time (a). Six typical bubble sheath electron motion trajectories in laboratory coordinates from 877 to 904 fs (b). (c) The evolution of the typical electron longitudinal velocity from 750 to 1000 fs and the marked interval is from 877 to 904 fs. (d) The electrons radiation power spectrum from 877 to 904 fs.

**Figure 3**

The variable of the bubble longitudinal radius $R_\parallel$ (circle black line) and transverse radius $R_\perp$ (star black line), and the emitted THz frequency (square blue line) with plasma density (a) and laser intensity (c), respectively. The relationship of THz frequency (black line) and the bubble radius aspect ratio $\eta = R_\parallel / R_\perp$ with plasma density (b) and laser intensity (d), respectively.

**Figure 4**

Snapshots of (a) the electron density and (b) the magnetic field when a relativistic laser ( $a_0 = 10, w_0 = 10$ and $\tau=21\text{fs}$ ) interacts with an underdense up-ramp plasma with $n_0 = 0.016 n_c$ at 1380 fs. (c) Temporal magnetic waveform and (d) corresponding Fourier spectrum of the backward radiation observed at a point ( $x = 180, y = 120$ ) located in vacuum.

**Figure 1**

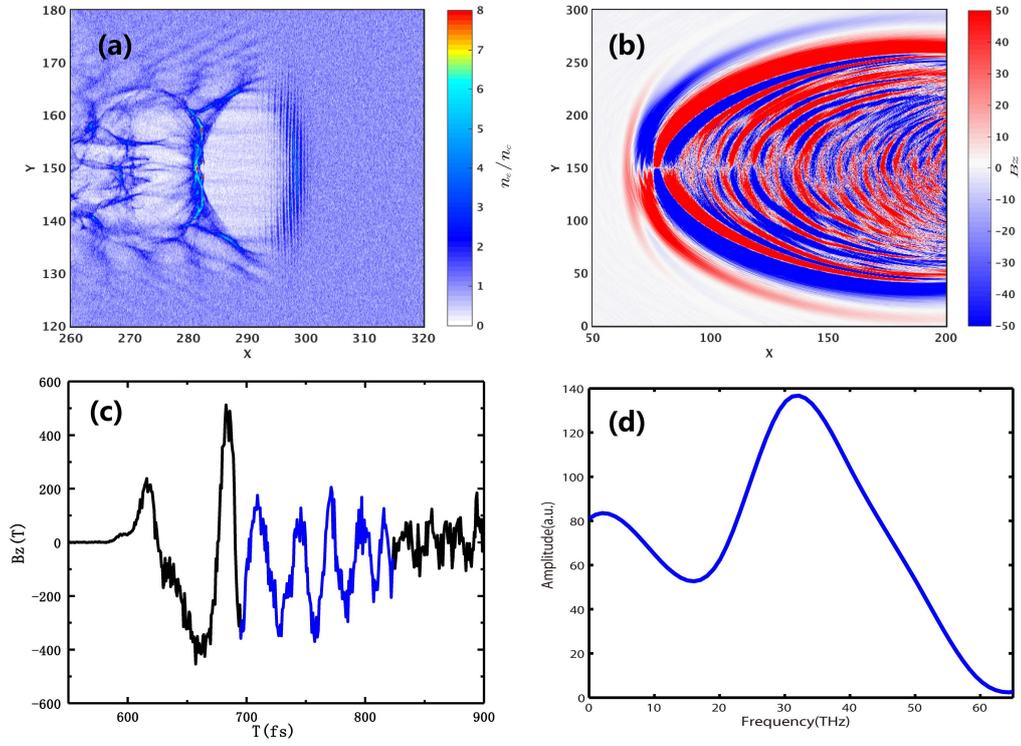

**Figure 2**

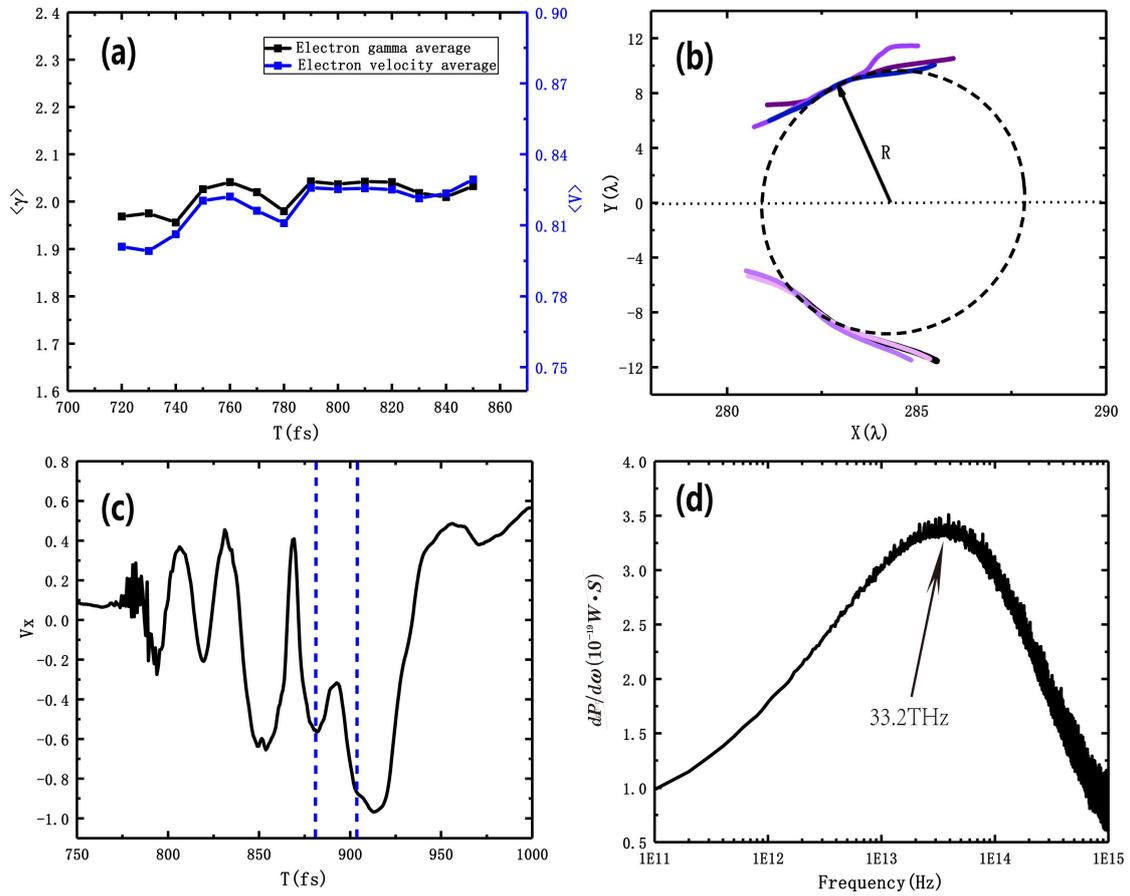

**Figure 3**

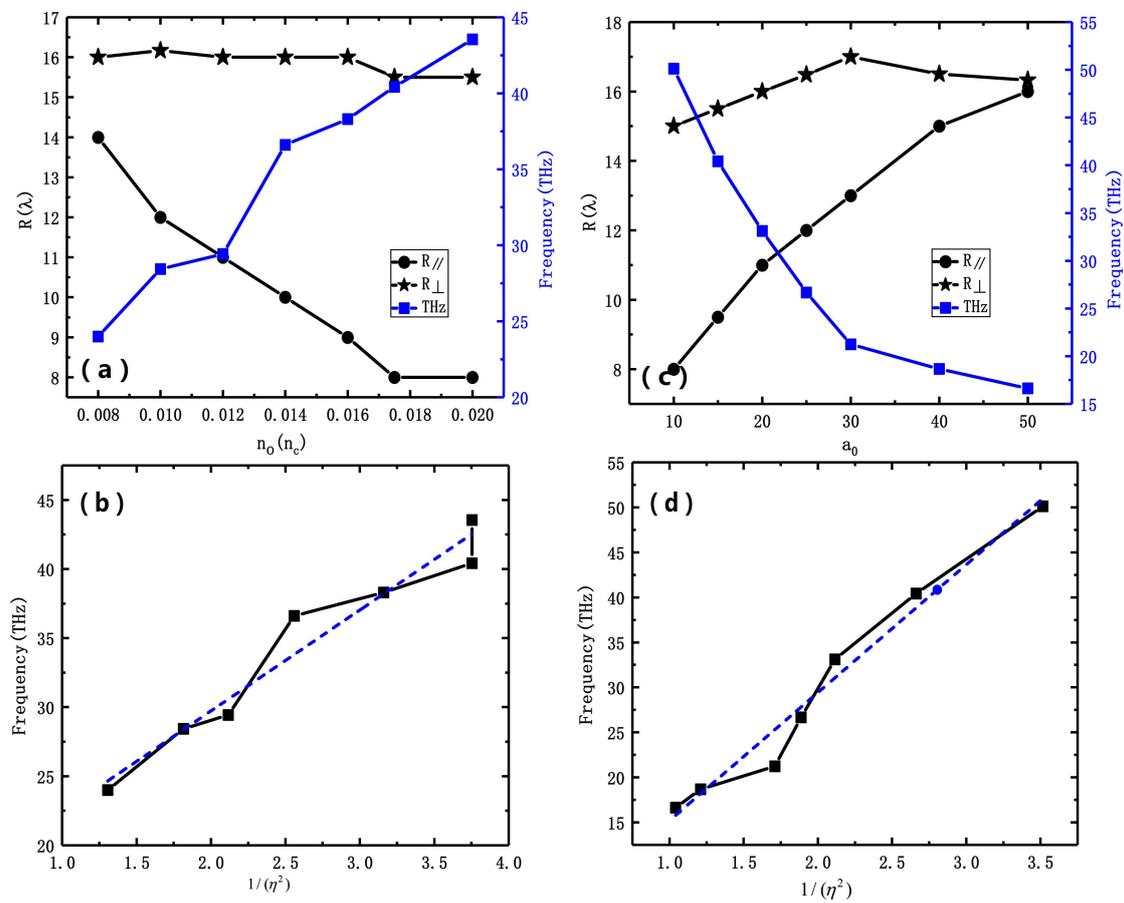

**Figure 4**

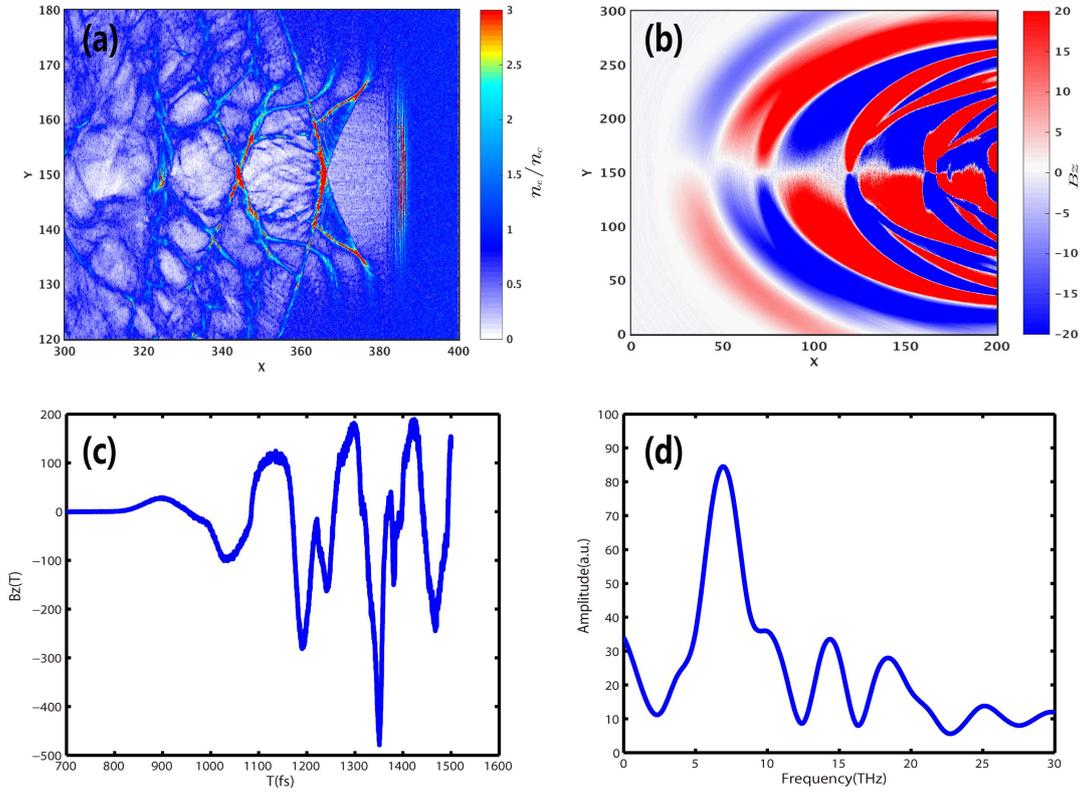